\documentclass[10pt,aps,pra,groupedaddress,nofootinbib]{revtex4-1}
\pdfoutput=1
\usepackage{graphicx}
\usepackage[english]{babel}
\usepackage{xcolor}
\usepackage{graphicx}
\begin{document}

%\preprint{}

\title{Transport of nano-objects in narrow channels: influence of Brownian diffusion, confinement and particle nature}


\author{Olivier Liot}
\altaffiliation[]{F\'ed\'eration FERMaT, INP Toulouse}
\altaffiliation[]{Presently at: Institut Lumi\`ere Mati\`ere, CNRS, Universit\'e Lyon 1, Villeurbanne, France}
\email[]{olivier.liot@laas.fr}
\author{Marius Socol}%
\author{L\'eo Garcia}
\author{Juliette Thi\'ery}
\author{Agathe Figarol}
\altaffiliation[]{Institut de Pharmacologie et de Biologie Structurale, Universit\'e de Toulouse, CNRS UMR 5089, Universit\'e Toulouse III - Paul Sabatier, F-31077 Toulouse, France}
\author{Anne-Fran\c oise Mingotaud}
\altaffiliation[]{Laboratoire des IMRCP, Universit\'e de Toulouse, CNRS UMR 5623, Universit\'e Toulouse III - Paul Sabatier, 118 route de Narbonne 31062 Toulouse Cedex 9, France}
\author{Pierre Joseph}
\email[]{olivier.liot@laas.fr,,pierre.joseph@laas.fr}
\affiliation{LAAS-CNRS, Universit\'e de Toulouse, CNRS UPR 8001, Toulouse, France}%Lines break automatically or can be forced with \\

%\thanks{}


\date{\today}

\begin{abstract}
This paper presents experimental results about transport of dilute suspensions of nano-objects in silicon-glass micrometric and sub-micrometric channels. Two kinds of objects are used: solid, rigid latex beads and {spherical} capsule-shaped, soft polymersomes. They are tracked using fluorescence microscopy. Three parameters are studied: confinement (ratio between particle diameter and channel depth), Brownian diffusion and particle nature. The aim of this work is to understand how these different parameters affect the transport of suspensions in narrow channels and to understand the different mechanisms at play. Concerning the solid beads we observe the appearance of two regimes, one where the experimental mean velocity is close to the expected one and another where this velocity is lower. This is directly related to {a competition between confinement, Brownian diffusion and advection}. These two regimes are shown to be linked to the homogeneity of particles distribution in the channel depth, which we experimentally deduce from velocity distributions. This inhomogeneity {appears} during the entrance process into the sub-micrometric channels{, as for hydrodynamic separation or deterministic lateral displacement}. Concerning the nature of the particles we observed a shift of transition towards the second regime likely due to the relationships between shear stress and polymersomes mechanical properties which could reduce the inhomogeneity imposed by the geometry of our device. 
\end{abstract}

%\keywords{}

\maketitle


\section{Introduction}

Transport of confined colloids in small channels is a key for analyzing many situations in biology (blood flow, flow cytometry, DNA analysis \cite{ranchon2016}), and flows in porous media \cite{bradford2008}  (chemical engineering with polymer processing, clogging \cite{wyss2006,dressaire2017}, separation \cite{wu2009}, geophysics i.e. fractured rocks \cite{zhang2012}). Owing to this ubiquity, the comprehension of the behaviour of advected simple particles or colloids in narrow channels has thus a lot of interests. It has been studied both theoretically \cite{staben2003} and experimentally \cite{staben2005}. The hydrodynamic interactions induced by a particle flowing close to a wall begin to be understood \cite{pasol2011}. These works led to a new flow metrology method based on particle transport in microchannels \cite{ranchon2015}. However in order to bear a close resemblance to natural or industrial systems, many points still need to be assessed and this field remains very active. For example, use of non-Newtonian fluids, with the appearance of transverse forces \cite{lu2017}, could have large industrial interest such as for DNA separation \cite{ranchon2016}. Effects of wall roughness \cite{charru2007,ranchon2018} or softness \cite{urzay2007,davies2017} are very important particularly in natural systems . The entrance geometry can also have substantial effects on particle distribution in a channel and subsequently their transport in a pore \cite{staben2005}. Furthermore transport of biomimetic vesicles and soft particles in microfluidic devices is an emerging topic with dramatic implications such as drug vectorization \cite{maeda2012} or comprehension of the transport of biological objects. For instance, a higher concentration of nano-objects transported by the blood inside the tumor tissues has been observed compared to healthy ones (Enhanced Permeability and Retention effect \cite{maeda2012}). It is hypothesized to be partly linked to margination \cite{kumar2012} (non-homogeneous and particle dependent, radial distribution of blood components, i.e. red blood cells, platelets, lymphocytes), in synergy with inter-cellular spaces (gaps) between endothelial cells (the major components of the blood vessel walls) larger within a tumour tissue than in normal ones. Such mechanism is a good way to target a tumour using encapsulated drugs. These capsules could be made with self-assembled block co-polymers, named polymersomes \cite{dionzou2016}. Behaviour of soft capsules or vesicles under flow is quite well documented \cite{lefebvre2007,vlahovska2009}. Nevertheless, the cross effects of Brownian diffusion, particle nature and confinement are still not well understood. Yet, in the case of sub-micrometric particles, these effects are crucial to explain the transport of nano-objects within the body, or in the context of porous media.

This paper presents some experimental results about transport of nanometric and sub-micrometric solid, rigid beads and {spherical} capsule-shaped, soft polymersomes in silicon-glass channels. We first investigate the coupled effect of confinement and Brownian motion on the transport of solid beads. We discuss the different results in light of inhomogeneous distribution of particles in the channels induced by entrance effects{, modulated by cross-effects between confinement, advection and Brownian diffusion}. Then, some results about polymersomes are detailed.

\section{Experimental method}

\subsection{Suspension of nano-objects}
\label{subsection:suspensions}
Two kinds of nano-objects are used: rigid, solid polystyrene beads, and soft, {spherical} capsule-shaped polymersomes. The polystyrene particles are commercial beads, with a density of 1.05 {g.mL$^{-1}$}. They are carboxylate-modified in surface, a fluorophore is loaded in volume and their zeta potential was measured between -49 and -69\,mV depending on the batch.  Their fabrication process allows a high mono-dispersity. The diameters and size distribution, provided by the manufacturer (measurements by Dynamic light Scattering, DLS \cite{stetefeld2016}), are compiled in table 1. Both values of the mean diameters and size distribution are consistent with our own DLS measurements. They are made on a Malvern NanoZS apparatus, using ``general purpose'' algorithm based on non-negative least squares (NNLS) analysis. The PolyDispersity Index (PDI, a normalized measurement of polydispersity) is of order 0.05, typical of monodisperse suspensions.

%Diameters $D_p$ of these nano-objects are compiled in table \ref{table:particles}. The diameters and uncertainties given by the manufacturer are consistent with our Dynamic Light Scattering (DLS) measurements \cite{stetefeld2016}Ï made with the Malvern NanoZS apparatus.

Polymersomes are self-assembled objects consisting in a double layer of block copolymer poly(ethylene glycol-b-methylmethacrylate) (PEO-PMMA 2000-5040\,g.mol$^{-1}$) with solvent inside. They are fabricated using the ``THF/MeOH cosolvent'' method described in \cite{dionzou2016} and their characterization is also described therein. Polymersomes are made fluorescent (emission wavelength: 515\,nm) by adding DiOC18 (3,3'-Dioctadecyloxacarbocyanine Perchlorate), obtained from Thermo Fisher, in low proportion (0.1\% w/w). Their zeta potential in the saline and pH conditions of the experiments, was equal to about -20\,mV. Almost no aggregation is visible. Mean diameter and polydispersity are not finely tunable using this process. Mean size and size distribution of the objects were measured using DLS. They are compiled in table \ref{table:particles}. These objects are more polydisperse than polystyrene beads. The polydispersity reported in Table 1 is extracted from the width of the size distribution determined by DLS NNLS analysis. In addition, the typical variability we obtained by repeating measurement of the mean polymersomes diameter on the same batch is much smaller than this size distribution, which we thus consider a good estimate of sample polydispersity. Even though the ``quality check'' of the algorithm indicates in some cases a deviation with the model used in the fitting of the correlation function, this value gives a first order estimation of the size distribution. Indeed, the PDI of the four batches of polymersomes are always in the range 0.12 to 0.25, typical values for which the algorithm used to determine the diameter and size distribution of objects is appropriate. {The mechanical characterization of such objects is challenging \cite{le_meins2011} and is the topic of dedicated studies. For example, using AFM measurements, Jaskiewicz \emph{et al.} \cite{jaskiewicz2012} measured the bending modulus and Young's modulus of PDMS-b-PMOXA polymersomes respectively to $7\pm5\times10^{-18}$\,J and $17\pm11$\,MPa. To our knowledge there is not specific study about PEO-PMMA polymersomes mechanical properties.}

These objects are separately dispersed in a solution buffered using Phosphate Buffered Saline \footnotemark[4]\footnotetext[4]{Composed of NaCl (137\,mM), KCl (2.7\,mM), Na$_2$HPO$_4$ (10\,mM) and K$_2$HPO$_4$ (1.8\,mM).} diluted 50 times {in water}. {Experiments are made in a temperature-regulated room (21$\pm0.5^\mathrm{o}$C) leading to a dynamic viscosity $\eta=0.98\pm0.01\,$mPa.s}. {The} ionic strength {reaches} $I=3$\,mM and $\mathrm{pH}=7.5$. The resulting Debye length is around 5\,nm. {With this short length compared to the particle and channel sizes, there is no ionic exclusion inside the channels.}  The concentration of the beads is fixed to $7.5\times10^8$\,mL$^{-1}$ to be suited for image analysis. It leads to low volume fractions from $\phi=3.9\times10^{-7}$ to $\phi=3.9\times10^{-4}$ depending on the diameter of the beads. We are not able to fix precisely the polymersomes concentration because of the polydispersity. Nevertheless the used concentration, chosen by optically determining the number of polymersomes per unit volume, is in the same order of magnitude as for beads in order to avoid interactions between objects.

\begin{table}[!h]
\caption{\label{table:particles}Mean diameter $d$ and polydispersity of the two kinds of nano-objects. For precisions about measurements technique, see the text.}
\footnotesize
\begin{tabular}{@{}cc}
\hline
Polystyrene beads&Polymersomes\\
\hline
$100\pm6$\,nm&$140\pm46$\,nm\\
$250\pm9$\,nm&$210\pm76$\,nm\\
$490\pm15$\,nm&$850\pm296$\,nm\\
$1000\pm25$\,nm&$1100\pm209$\,nm\\
\hline
\end{tabular}\\

\end{table}

\subsection{Microfluidic device and observation}

\begin{figure}[h]
\begin{center}
\includegraphics[width=0.40\textwidth,angle=-90]{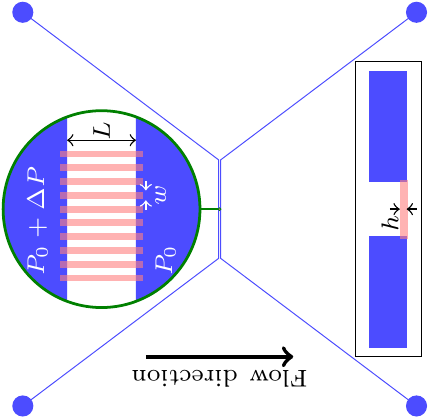}
\caption{Top view of the chip design used for the experiments. The microchannels are blue-colored; the nanoslits, visible in the zoom (green circle) are in red. The four blue circles represent the supply wells. Inset: side view of the chip.}
\label{puce}
\end{center}
\end{figure}

\begin{figure}[h]
\begin{center}
\includegraphics[width=0.40\textwidth]{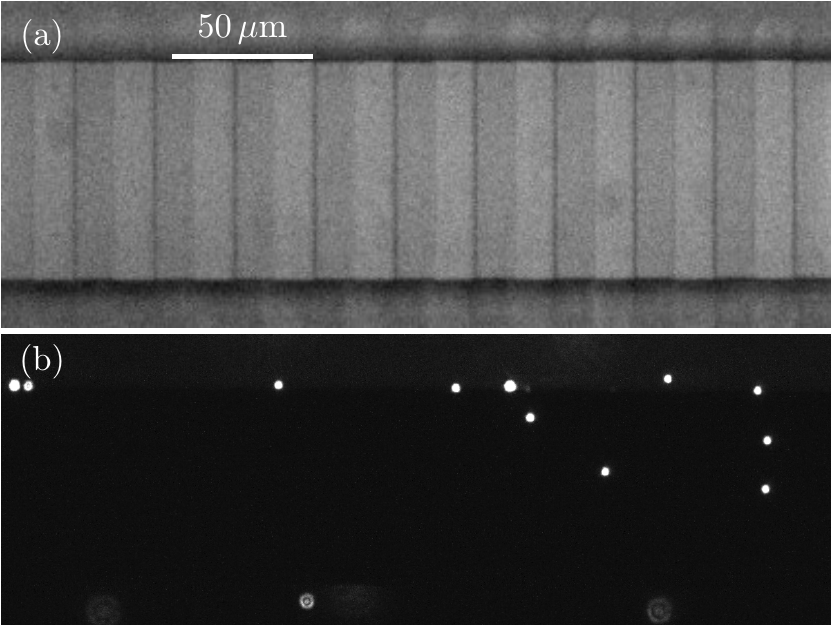}
\caption{(a) Picture in white light of the nanoslits. (b) Fluorescence image of particles flowing in the nanoslits. Scale bar is equal for both pictures.}
\label{photos}
\end{center}
\end{figure}

We use a versatile model system made of nanoslits etched in silicon (zeta potential around -30\,mV at $\mathrm{pH}=7.5$ \cite{kirby2004}) and covered with a 170 $\mu$m-thick borosilicate layer (root mean square roughness is inferior to 1\,nm on 1\,$\mu$m$^2$). Two microchannels are connected, from their bottom corner, by ten nanoslits (width $w=10\,\mu$m, length $L=50\,\mu$m, period $\delta=20\,\mu$m). Chips with different nanoslits depth have been used: $h=330,~830,~980,~1300,~1650~\mathrm{and}~3390$\,nm. The uncertainty on the nanoslits depth measured by mechanical profilometry and calibrated AFM is about 1\%. {With the process used to fabricate these chips, the bonding does not affect the nanoslits depth \cite{naillon2016}.} Figure \ref{puce} sketches the chip design. Figure \ref{photos}\,(a) shows a bright field microscopy (reflection mode) picture of the nanoslits.

The flow in the nanoslits is pressure-driven (pressure drop $\Delta P$) using a pressure controller \emph{Fluigent MFCS} whose sensitivity is about 0.02\,mbar. The particles are observed by fluorescence microscopy using a 40$\times$ magnification objective with 1.4 numerical aperture. Acquisitions are registered using a sCMOS camera with a sampling between 100 and 400 fps and an exposure time between 0.5 and 2\,ms, depending on the mean velocity of the particles. Figure \ref{photos}\,(b) shows a fluorescence picture of particles flowing in the nanoslits. Particles are tracked using a home-made script combining \emph{Python} and \emph{Matlab} routines. {Velocities are deduced from trajectories along the whole length of the nanoslits.} In order to have high statistics, each run lasts 1 to 2 minutes making us able to track up to more than $10^4$ particles giving more than $10^5$ velocity events.

\subsection{Model of transport in narrow channels}
\label{subsection:model}
To describe the transportation of particles in a confined channel, several mechanisms are involved. First, a particle does not {experience} an uniform velocity field on its surface. Consequently its velocity does not correspond to the fluid velocity in its center. Its actual velocity can be computed, in the case of a dilute suspension, using Fax\'en law \cite{chen2000}: 

\begin{equation}
\overrightarrow{V_p} = \overrightarrow{v} + \frac{d^2}{24}{\bigtriangleup} \overrightarrow{v},
\end{equation}

\noindent where $\overrightarrow{V_p}$ is the particle velocity, $\overrightarrow{v}$ is the fluid velocity at the center of the particle {of diameter $d$} and $\overrightarrow{\bigtriangleup}$ represents the Laplacian operator. It corresponds to the integral of the fluid velocity over the full surface of the particle facing the flow. 

Because of the aspect ratio of the nanoslits in some cases we cannot reduce the flow to a Poiseuille parabolic profile between two infinite plates. Instead of that we consider a laminar flow inside a rectangular pipe. At the first order we use the velocity field along the nanoslit length  (coordinate $x$) \cite{bruus2007}: 

\begin{equation}
v_x(y,z)=\frac{4h^2\Delta P}{\pi^3\eta L}\left(1-\frac{\cosh\left(\pi\frac{y}{h}\right)}{\cosh\left(\pi\frac{w}{2\,h}\right)}\right)\sin\left(\pi\frac{z}{h}\right),
\label{eq:vitesse}
\end{equation}

\noindent where {$z\in [0,h]$} is the coordinate in the depth direction and {$y\in [-w/2,w/2]$} in the width one. $\eta$ represents the dynamic viscosity of the fluid. {We remind that $L$ is the length of the nanoslits, $w$ their width, $h$ their depth and $\Delta P$ the imposed pressure drop.} {This expression is obtained with the same hypothesis as the Poiseuille's law, especially the no-slip condition at the walls.} The error {on the flow rate is lower than 0.2\% in the range of $h/w\in[0.033,0.34]$ we explore \cite{bruus2007}. We can thus consider that equation \ref{eq:vitesse} is robust enough to describe the flow in our nanoslits.}

However when a particle is transported by a laminar flow, it disturbs streamlines and excites long-range flows. Retroactively the particle moves in response to fluid motion. Close to a wall it leads to a phenomenon called Hydrodynamic Interactions (HI) \cite{bhattacharya2005}. Few years ago, Pasol \emph{et al.} \cite{pasol2011} proposed a way to compute these hydrodynamic interactions acting on a particle transported between two infinite plates. 

In order to have the expected mean velocity of a particle inside a channel, the velocity $\overrightarrow{V_p}$ {(including Fax\'en's law) is} corrected by the HI using the work of ref. \cite{pasol2011}. {Since we work with a dilute suspension (typically one particle in each nanoslit at the same time), we neglect the multi-particle interactions.} {The explicit formula, obtained from combined analytical and numerical approaches is particularly complex. The reader can see the detail in section 4 of the last reference.} {Then the particle velocity} is integrated on the whole channel excluding zones where the center of the particle would be closer to the wall than its own radius: it cannot inter-penetrate the walls. {Although the geometry of our device could lead to effects similar to deterministic lateral displacement \cite{mcgrath2014} or hydrodynamic separation \cite{yamada2004}, w}e make the underlying hypothesis that the particles are homogeneously distributed in the nanoslits depth. We define as $V_t$ the resulting mean velocity in the nanoslit direction. This predicted velocity is proportional to the pressure drop: $V_t=\alpha_t \Delta P$. 

\section{Solid nano-objects: influence of geometry and Brownian diffusion}

\subsection{Mean velocity: two regimes}

\begin{figure}
\begin{center}
\includegraphics[width=0.40\textwidth]{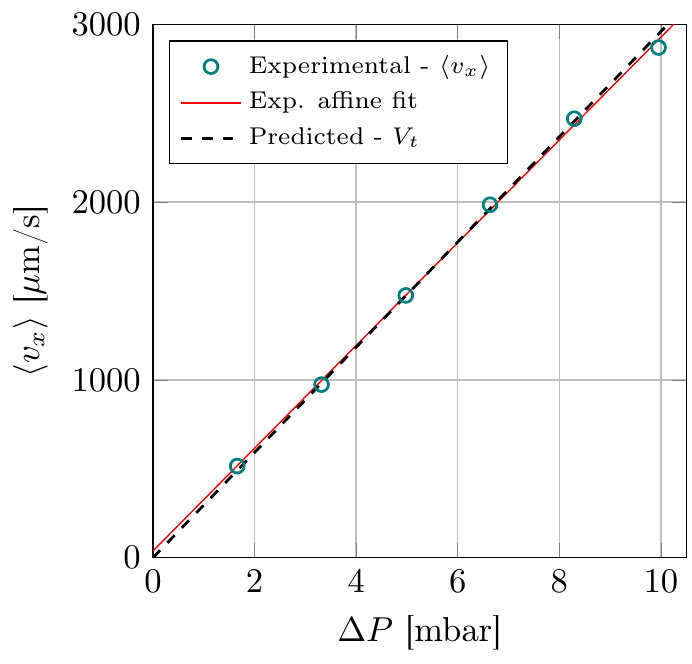}
\caption{Experimental and predicted mean velocity of a bead ($d=1000$\,nm) flowing in a nanoslit ($h=1300$\,nm) as a function of the pressure drop. An affine fit of the experimental points is added. Error bars are smaller than the points extension.}
\label{comparaison}
\end{center}
\end{figure}

We first present results for solid beads. For each configuration (bead diameter and nanoslits depth), the experimental mean velocity is measured for 3 to 8 different pressure drops. Let us recall that averages are made on around $10^5$ velocity events. Figure \ref{comparaison} shows the experimental mean velocity as a function of the pressure drop for beads with diameter $d=1000$\,nm flowing in nanoslits with depth $h=1300$\,nm. The predicted velocity $V_t$ is also plotted. Experimental points are fitted using an affine law to correct possible systematic bias in the effective pressure drop applied to nanoslits: $\langle v_x\rangle = \alpha_{exp}\Delta P + \Delta P_0$.

 To compare experimental and predicted velocities the ratio $\beta=\alpha_{exp}/\alpha_t$ is computed. The error bars are estimated by uncertainty propagation. We take into account the uncertainty on the nanoslits height (1\%), the bead diameter (see table \ref{table:particles}) and the uncertainty due to 95\% confidence bound on the experimental data fit.
 
 % For clarity of data presentation, we arbitrarily divide the objects in two groups: highly Brownian ($d\leq 250$\,nm) and hardly Brownian ($d\geq 490$\,nm). Furthermore the confinement is defined as:
 
 \begin{equation}
 r=\frac{d}{h}.
 \end{equation}
 
 \begin{figure}[h]
\begin{center}
\includegraphics[width=0.49\textwidth]{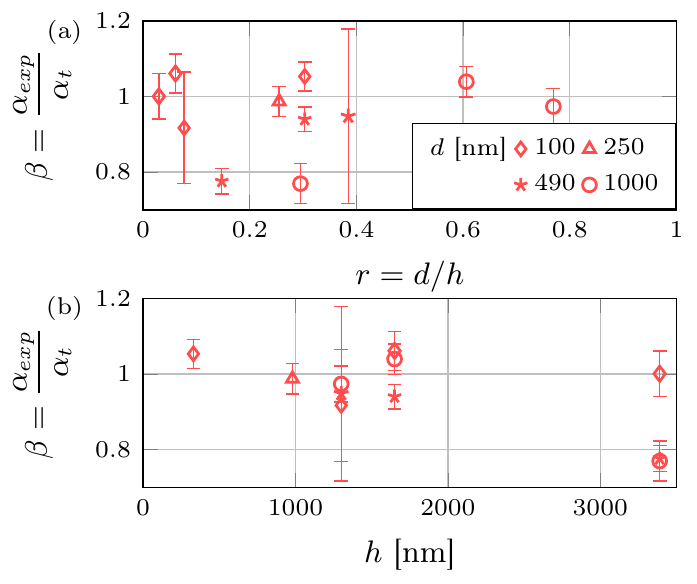}
\caption{Ratio for {rigid beads} between experimental and predicted velocity proportionality coefficients with $\Delta P$ as a function of (a) the confinement and (b) the nanoslits depth. See text for details about error bars. {The legend is valid for both plots.}}
\label{meanv_b_fit}
\end{center}
\end{figure}
 
 Figure \ref{meanv_b_fit} shows the ratio $\beta$ for {rigid} beads versus (a) the confinement and (b) the nanoslits depth. {For most of} the confinement values or the nanoslits depth, the experimental mean velocity is close to the predicted one ($\beta$ close to 1) within the experimental errors{, except for two configurations}. {The beads of diameter $d=490$\,nm and 1000\,nm in the nanochannels of depth $h=3390$\,nm reveal} a velocity more than 20\% lower than the expected one. {The} agreement with predictions is quite remarkable for such high confinements (up to around 0.8). The hydrodynamic model, including only velocity field averaging and  hydrodynamic interactions, has thus a quite extended range of validity.  We hypothesize that {the difference between the first group ($\beta\sim1$) and the second one ($\beta<0.8$)} could be related to {a change} of the distribution of the beads in the nanoslit depth.

%\begin{figure}[h]
%\begin{center}
%\includegraphics[width=0.49\textwidth]{Figure5}
%\caption{Ratio for rigid beads between experimental and predicted velocity proportionality coefficients with $\Delta P$ as a function of (a) the confinement and (b) the nanoslits depth. See text for details about error bars. The legend is valid for both plots.}
%\label{meanv_nb_fit}
%\end{center}
%\end{figure}
% The large error bar is due to an experiment with few pressure drops leading to a large uncertainty in the affine fit.

% Figure \ref{meanv_nb_fit} shows the ratio $\beta$ for hardly Brownian beads versus (a) the confinement $r$ and (b) the nanoslits depth $h$. We notice that the behaviour is quite different. Two groups appear: beads in nanoslits of $h=1300$ and 1650\,nm and beads inside the deepest nanoslits ($h=3390$\,nm). The first group has a mean velocity close to the predicted one ($\beta\approx 1$). This agreement with predictions is quite remarkable for such high confinements (up to around 0.8). The hydrodynamic model, including only velocity field averaging and  hydrodynamic interactions, has thus a quite extended range of validity. The second group has a velocity more than 20\% lower than the expected one. We hypothesize that this could be related to violation of the homogeneous distribution of the beads in the nanoslit depth. 
 
 Figure \ref{meanv_b_fit}\,(a) reveals {also} a trend which is similar for {the two biggest} beads {($d=490$ and 1000\,nm)}. The ratio $\beta$ increases with the confinement $r$, but with a shift. The weak Brownian diffusion of these beads could eventually  permit them to have a quite homogeneous distribution in the nanoslits if the confinement is high enough (and consequently the distance to the walls/center is short). The shift in confinement observed between $d=490$\,nm and $1000$\,nm could be due to the higher Brownian diffusivity of the $d=490$\,nm beads: at a given confinement, the smaller are the beads, the easier they will reach the center or the walls by diffusion. Consequently, beads will statistically explore the whole accessible velocity range. This inhomogeneous repartition should also exist for {smaller and so more} Brownian beads, but their strong diffusivity enables a homogeneous distribution in the nanoslits.
 
{Inhomogeneous particle distributions have already been observed in confined channels. They can have different origins such as transverse lift during the particle travel or steric/entrance effects.}

{Dersoir \cite{dersoir2015-1} (p. 75) observed} an inhomogeneous distribution of beads flowing in microchannels for a confinement $r=0.29$. {T}he centers of the beads had a bimodal repartition in the channel depth{, with peaks at about one particle diameter}. Thus, the high-velocity streamlines at the center of the channels were rarely followed by beads. The origin is probably a transverse lift of the beads during their transport. Decades ago, some experimental and {theoretical} studies showed that neutrally buoyant beads have a transverse lift velocity in a Poiseuille flow even for a Reynolds number very small compared to 1 \cite{vasseur1976,schonberg1989}. The Reynolds number compares inertia to viscosity: 

\begin{equation}
Re=\frac{\rho U_mL}{\eta},
\end{equation}

\noindent where $\rho$ represents the density of the fluid, $U_m$ the maximal velocity of the flow and $L$ a typical length of the system. In our case, the Reynolds number based on the bead diameter (1\,$\mu$m) reaches about $10^{-3}$ for a maximal typical velocity $U_m$ of 1\,mm/s. Using the Vasseur \& Cox work \cite{vasseur1976} (figure 9) we can estimate the lift velocity of a bead in the nanoslits to $v_l\sim 10^{-4}U_m$. {In our nanoslits,} the typical drift length is {only} few nanometers. This is consistent with the absence of experimental evidence of significant drift in the nanoslits: the mean velocity does not depend on the location in the nanoslits along the flow (not shown here). It thus means that the inhomogeneous distribution occurs prior to or at the vicinity of the nanoslits entrance.

If an inhomogeneity of bead distribution is observed in the inlet microchannel, it should lead to a similar inhomogeneity in the nanoslits \cite{dersoir2015-1,dersoir2017}. Because the velocity in reservoirs from where the suspension is injected in the chip is almost null, we assume beads are distributed homogeneously by Brownian diffusion (even for the biggest objects). The length of an inlet microchannel is about 5900\,$\mu$m from the well (blue circle figure \ref{puce}) to the nanoslits entrance. For the microchannels whose depth is 23\,$\mu$m, the typical maximal velocity is 0.05\,mm/s. Such a velocity and a lower confinement than in the nanoslits will lead to a negligible drift in the microchannel.

%A possible mechanism could be imagined from work of Guglielmini \emph{et al.} \cite{guglielmini2011}. They numerically showed that for $Re=10^{-3}$ some three-dimensional secondary flows appear at the vicinity of a 90\textsuperscript{o} corner in a rectangular pipe. Such a phenomenon, coupled to the finite size of the beads could lead to a ``selection'' of streamlines by the objects. Nevertheless it does not propose a quite robust explanation of the inhomogeneous distribution of the beads in the nanoslits.

\begin{figure}[h]
\begin{center}
\includegraphics[width=0.48\textwidth]{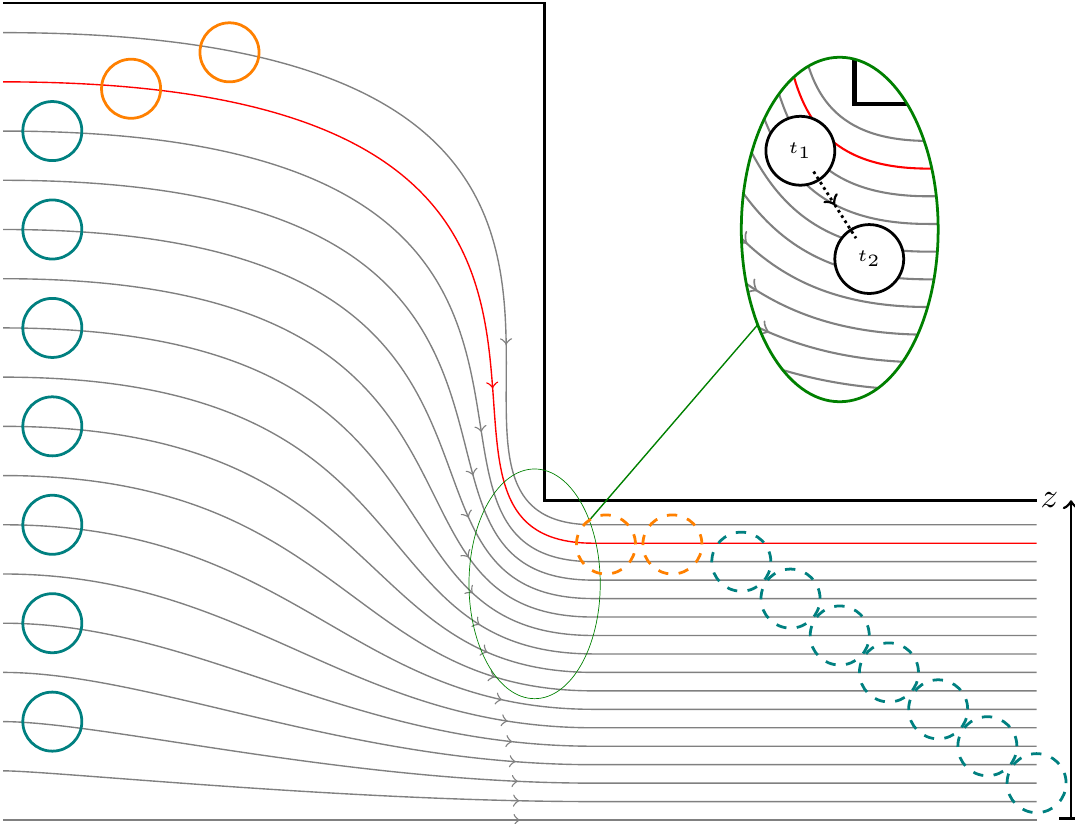}
\caption{Sketch (not to scale) of the supposed nanoslit entrance concentration mechanism. Gray/red lines represent streamlines. Whereas teal beads will remain on the same streamline before and after the entrance in the nanoslit, orange ones, above or on the red streamline, will roughly be on the same streamline in the nanoslit, pushed by the corner similarly to hydrodynamic filtration \cite{yamada2004,fouet2016}. The zoom shows a possible displacement (due to advection and diffusion) of a bead at the entrance of the pore between time $t_1$ just before entering and time $t_2$ just after entering.}
\label{entrance}
\end{center}
\end{figure}

The only remaining hypothesis is an inhomogeneous distribution in the nanoslits due to the entrance geometry{, similarly to deterministic lateral displacement \cite{mcgrath2014} or hydrodynamic separation \cite{wu2009}}. The step which connects microchannel and nanoslits (see inset figure \ref{puce}) could actually lead to {a similar} behaviour when beads cross this interface. If we neglect diffusion, beads whose mass center is on a streamline in the microchannel will stay on this streamline inside the nanoslit, except if the streamline distance to the top wall is lower than beads radius, in an analogous way as for hydrodynamic filtration \cite{wu2009,yamada2004,fouet2016}. This specific geometry and the finite size of objects let beads cross the streamlines when they are tightened at the entrance of the nanoslits \cite{dersoir2015-1,dersoir2017}. First, because of streamlines tightening in the nanoslits {compared to the microchannel}, some beads close to the top wall in the microchannel won't be able to remain on the same streamline in the nanoslit because of the exclusion zone (no inter-penetration wall/beads). {In addition}, this tightening could be more important close to the corner of the step and it could amplify this phenomenon {by affecting more particles}. Consequently, beads moving in a specific bandwidth close to the top wall of the microchannel will be concentrated at the vicinity of the top wall of the nanoslit. Figure \ref{entrance} sketches the situation. It will lead to a bead concentration heterogeneity with more beads close to the top wall and a non-flat $z$ position distribution of the beads in the nanoslits depth. The diffusion during the entrance could restore bead distribution homogeneity.

{Note that for $\beta <1$ we do not observe effect of the lateral diffusion along the $z$ axis during the particle travel in the nanoslits. When $\beta\sim 1$, since particles distribution is homogeneous from the entrance of the nanoslits, possible lateral diffusion along the $z$ axis (see figure \ref{entrance}) cannot affect velocity statistics.} Eventually the combination of Brownian motion and confinement could be the good point of view to understand the different behaviours observed on the experimental mean velocity of beads in confined nanoslits.  We thus construct an ``entrance P\'eclet number'' $Pe_{entrance}$ to compare the typical time of advection and diffusion at the entrance of the pore. To simplify, we assume a bead at the entrance of a nanoslit at time $t_1$, but entirely in the microchannel. Zoom of figure \ref{entrance} draws this situation. The typical time for the bead to diffuse from the top wall minus bead radius altitude to the center, which is the length necessary to homogenize the bead distribution by diffusion, is: 

\begin{equation}
t_{diff}=\frac{1}{2D}\left(\frac{h-d}{2}\right)^2,
\end{equation}

\noindent where $D=k_b T/3\,\pi\eta d\sim10^{-12}$\,m$^2$.s$^{-1}$ is the Brownian diffusion coefficient of the bead at a temperature $T$ ($k_b$ represents the Boltzmann constant). During the entrance process, between the time a bead starts entering in the nanoslit and the time $t_2$ it is completely inside, the bead typically moves on a distance $d$ at a velocity $V_t$. Thus the typical advection time at the entrance of the pore is:

\begin{equation}
t_{adv}=\frac{d}{V_t}.
\end{equation}

\noindent The P\'eclet number at the entrance of the nanoslit compares these two characteristic times: 

\begin{equation}
Pe_{entrance}=\frac{t_{diff}}{t_{adv}}=\frac{3\pi\eta}{8k_b T}V_th^2(1-r)^2.
\label{eq:Pe}
\end{equation}

\begin{figure}[h]
\begin{center}
\includegraphics[width=0.49\textwidth]{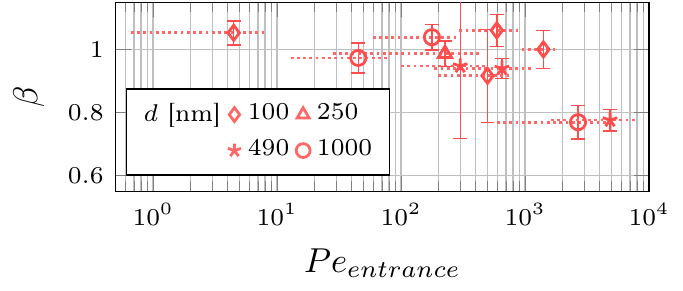}
\caption{Ratio between experimental and predicted velocity proportionality coefficients with $\Delta P$ as a function of the entrance P\'eclet number. The dashed lines show the extension of $Pe_{entrance}$ corresponding to the different used $\Delta P$, the point is chosen to be at the center of the P\'eclet range.}
\label{beta_pe}
\end{center}
\end{figure}

Figure \ref{beta_pe} shows the experimental ratio $\beta$ as a function of this entrance P\'eclet number. Since this number directly depends on the mean object velocity $V_t$, we plot the range of P\'eclet for each experiment. Two regimes related to the bead distribution in the nanoslits are visible: the ``homogeneous'' and ``inhomogeneous'' regimes. The two points corresponding to beads in the deepest nanoslits, with $\beta<1$ are very close {to} each other whereas other points representing different beads in the same nanoslits are not. Moreover they correspond to a high entrance P\'eclet number when diffusion and confinement are much too low to have a homogeneous distribution. It confirms that for these two configurations ($h=3390$\,nm; $d=490$ and 1000\,nm), the bead transport obeys to the same mechanisms and belongs to the ``inhomogeneous'' regime. Then we notice that points {corresponding to large and small beads} are mixed approximatively between $Pe_{entrance}=10^2$ and $10^3$. This similar behaviour is due to a cross effect between diffusion and confinement. Either beads are small enough to homogenize beads at the nanoslits entrance whatever the confinement, or if the Brownian diffusion is low, the confinement is high enough to have homogenization anyway.  Additionally, the point at $Pe_{entrance}\approx4$ combines high Brownian diffusion ($d=100$\,nm) and quite high confinement ($r=0.3$). All these points belong to the ``homogeneous'' regime. Even if this could be considered as a design-dependent effect because it is an entrance effect, the mechanism is quite general because nanochannels are always connected to larger channels or reservoirs. This general  representation using a P\'eclet number highlights the importance to consider both confinement and Brownian diffusion to understand the transport of beads in narrow channels. 

%They have similar behaviour for different reasons: either the Brownian diffusion is strong enough to mix the beads at the entrance and prevent from an inhomogeneous distribution; or the confinement is high enough to allow, even for weak Brownian diffusion, a homogeneous distribution at the entrance of the nanoslits.

\begin{figure}[h!]
\begin{center}
\includegraphics[width=0.45\textwidth]{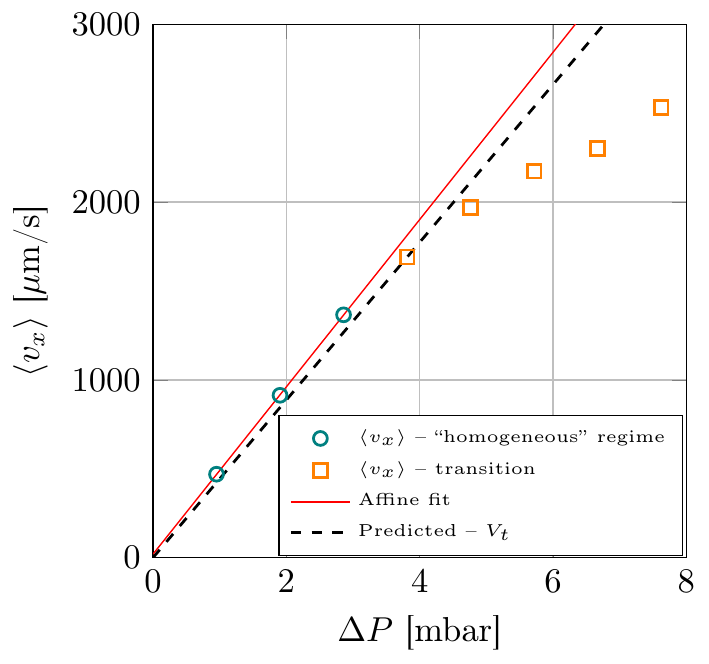}
\caption{Experimental mean velocity of beads ($d=100$\,nm) flowing in a nanoslit ($h=1650$\,nm) as a function of the pressure drop. Points are separated in two groups: ``homogeneous'' regime and transition. The affine fit {to compute $\beta$ in figures \ref{meanv_b_fit} and \ref{beta_pe}} is made on the points in the ``homogeneous'' regime.}
\label{transition}
\end{center}
\end{figure}

 The existence of these two regimes related to the distribution of the beads in the nanoslit depth should let appear a transition in certain conditions on confinement, diffusion and object velocity. {An other way to change the entrance P\'eclet number consists in varying the advection velocity, or pressure drop. The results presented in figure \ref{meanv_b_fit} and \ref{beta_pe} show two clear, well separated regimes.} Actually, experiments made with beads of $d=100$\,nm reveal an interesting behaviour. The mean velocity as a function of the pressure drop for a bead of $d=100$\,nm transported in nanoslits of depth $h=1650$\,nm is plotted on figure \ref{transition}. We separate {on this figure} the points in two groups: the ``homogeneous'' regime where the ratio $\beta\approx 1$ and the transitional state when $\langle v_x\rangle$ deviates from $V_t$. For $\Delta P>3$\,mbar, the experimental mean velocity departs from the predicted velocity, transiting from the ``homogeneous'' towards the ``inhomogeneous'' regime. Such a {non linear} behaviour is also observed for beads of diameter $d=100$\,nm transported in nanoslit of depth $h=1300$ and 3390\,nm. {On the contrary, mean experimental velocities in other configurations remain linear  in the explored $\Delta P$ range.} We point out that {for the beads with diameter $d=100$\,nm}, the fit used to compute $\beta$ in figures \ref{meanv_b_fit} and \ref{beta_pe} is made on the points belonging to the ``homogeneous'' regime{, as described on figure \ref{transition}}.

Such a behaviour reveals that when the advection velocity increases, the diffusion is not able to homogenize the beads at the entrance of the nanoslits anymore. In term of entrance P\'eclet number, the three points corresponding to these experiments (diamonds in figure \ref{beta_pe}) have a $Pe_{entrance}$ a bit lower than the two points in the ``inhomogeneous'' regime. In the case plotted figure \ref{transition}, the P\'eclet number {is larger than} 2000 for $\Delta P\geq7.5$\,mbar. This is quite consistent with the P\'eclet numbers observed for the points in the ``inhomogeneous'' regime. {Furthermore, competition between advection and lateral diffusion (along the $z$ axis) during the particles travel inside the nanoslits could affect the mean experimental velocity. In the last presented case, with so Brownian beads, we observed that lateral diffusion starts to affect the velocity statistics along the nanoslits when the initial distribution is inhomogeneous ($\Delta P>4$, data not shown). In this case, the mean velocity is not constant along the nanoslits. For the lower pressure drop, since the initial distribution is homogeneous ($\beta\sim1$), lateral diffusion cannot affect the transport in the nanoslits. Notwithstanding we do not use these points to compute figures \ref{meanv_b_fit} and \ref{beta_pe} so it does not affect our results.}

%Nevertheless the difference between the inlet microchannel depth (23\,$\mu$m) and the nanoslits one 
 
\subsection{Confirmation of {both homogeneous and inhomogeneous distributions}}

To go further in the analysis of these two regimes, we look into the velocity and position distributions. 

\begin{figure}[h!]
\begin{center}
\includegraphics[width=0.45\textwidth]{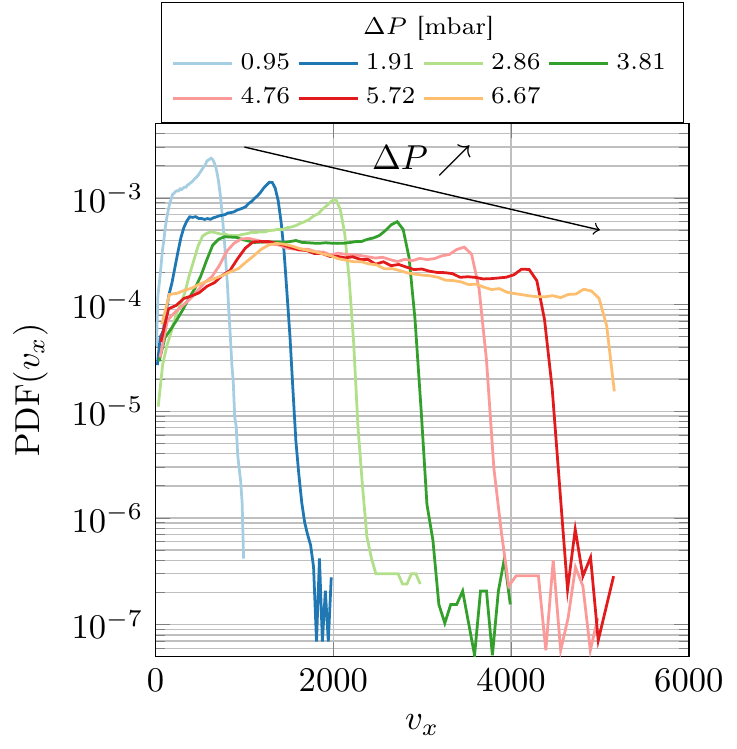}
\caption{Probability density functions (PDF) of the longitudinal velocity $v_x$. Beads have a diameter $d=100$\,nm and are transported in nanoslits of depth $h=1650$\,nm. Because of lack of statistics, the case $\Delta P=7.62$\,mbar is not shown.}
\label{pdf_transition}
\end{center}
\end{figure}

Figure \ref{pdf_transition} presents the evolution of the Probability Density Functions (PDF) of the longitudinal velocity $\langle v_x\rangle$ for beads of diameter $d=100$\,nm transported in nanoslits of depth $h=1650$\,nm. {It corresponds to the mean velocities presented on figure \ref{transition}.} We observe a dramatic change of the PDF shape. For the lower $\Delta P$ a peak of most probable velocities is clearly higher than the low velocity tail of the distributions. These are the same distributions as the ones from Ranchon \emph{et al.} \cite{ranchon2015} made in very similar conditions. When $\Delta P$ is increased the importance of the tail grows up to become higher than the peak. This is consistent with the observations made on figure \ref{transition}: the high velocities become less important than the low ones leading to a reduction of the mean velocity compared to the predicted one. {On figures \ref{meanv_b_fit} and \ref{beta_pe}, beads of diameter $d=100\,$nm in nanoslits of depth $h=1300$, 1650 and 3390\,nm are in the regime $\beta\sim1$ because we selected only the low $\Delta P$ velocities to compute these points. But actually these points are very close to the transition, which appears when the advection is increased.}

\begin{figure}[h!]
\begin{center}
\includegraphics[width=0.45\textwidth]{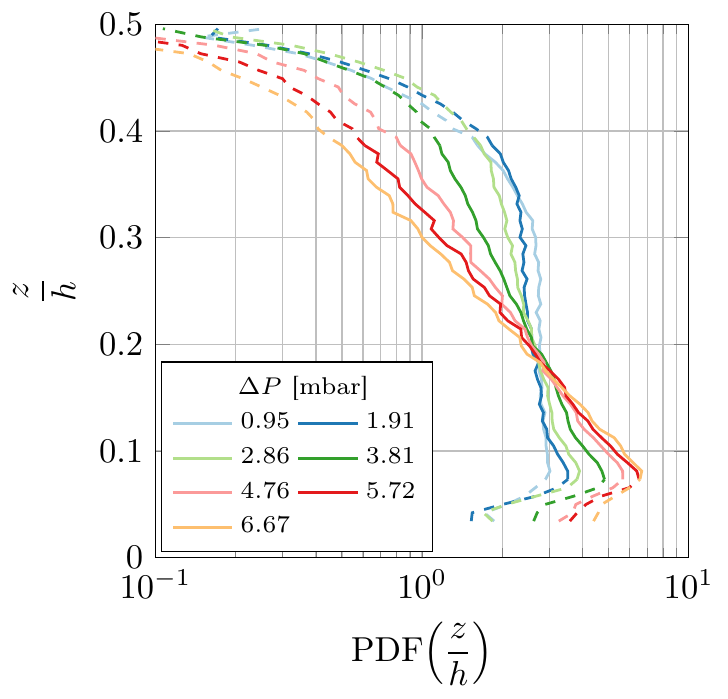}
\caption{PDF of the position $z$ of the beads in half of the nanoslits depth. Dashed parts correspond to bias-affected positions (see text for details). The center of the nanoslits correspond to $z/h=0.5$. Beads have a diameter $d=100$\,nm and are transported in nanoslits of depth $h=1650$\,nm. Because of lack of statistics, the case $\Delta P=7.62$\,mbar is not shown.}
\label{pdf_posz_transition}
\end{center}
\end{figure}

As we proposed in the previous subsection, such an evolution could be related to a transition from a regime where the beads are distributed homogeneously in the nanoslits depth towards a regime where it is not the case anymore. {Since the microscope focal plane is perpendicular to the nanoslit depth axis, we do not have a direct access to the particle depth position.} {Moreover, methods using the defocused image of particles (Point Spread Function) would be useless here. The velocity of the particles leads to a noisy pattern, and the depth of field of the objective is not small enough with respect to the typical nanoslits depth.} {Nevertheless,} the position distribution can be computed assuming a Poiseuille flow in the nanoslits. Using the model described at subsection \ref{subsection:model} and neglecting in a first phase Brownian diffusion, we are able to attribute to an experimental velocity value its vertical position $z$ in the nanoslit. Nevertheless the symmetry of the flow does not permit to distinguish two symmetric positions on either side of the nanoslit center. The particle position PDF can be computed only in a half of the nanoslits depth. Thus, we cannot observe with this method a potential asymmetry of the $z$ PDF.  Furthermore, because of the Brownian motion, the actual velocity of the bead is not the one it should have at its vertical position $z$ in the nanoslit. This introduces a bias at the two extremes of the accessible velocity range whereas in its center the effect is auto-compensated. In this central zone and because the velocity distribution due to the Brownian motion is symmetric, if a velocity is overestimated due to the Brownian motion, it will be statistically compensated by an underestimated velocity. The results are plotted on figure \ref{pdf_posz_transition}. The walls correspond to $z/h=0$ and the center of the nanoslit to $z/h=0.5$. The dashed parts roughly correspond to zones where the PDF is affected by the bias mentioned above. In the bias-unaffected zones we observe a clear transition from a flat distribution {for pressure drops up to $\Delta P=2.86$ towards} a higher concentration close to the walls for higher pressure drops. This is a net evidence of a transition to an ``inhomogeneous'' regime corresponding to a drift of the experimental mean velocity compared to the predicted one.

\begin{figure}[h!]
\begin{center}
\includegraphics[width=0.45\textwidth]{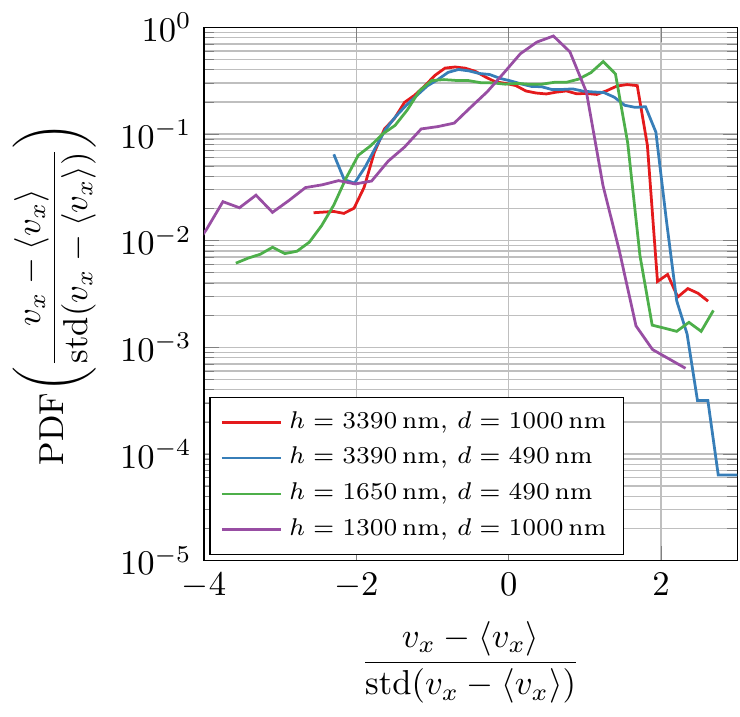}
\caption{PDF of the longitudinal velocity $v_x$ in different configurations for {the two largest} beads.}
\label{pdf_nb}
\end{center}
\end{figure}

\begin{figure}[h!]
\begin{center}
\includegraphics[width=0.45\textwidth]{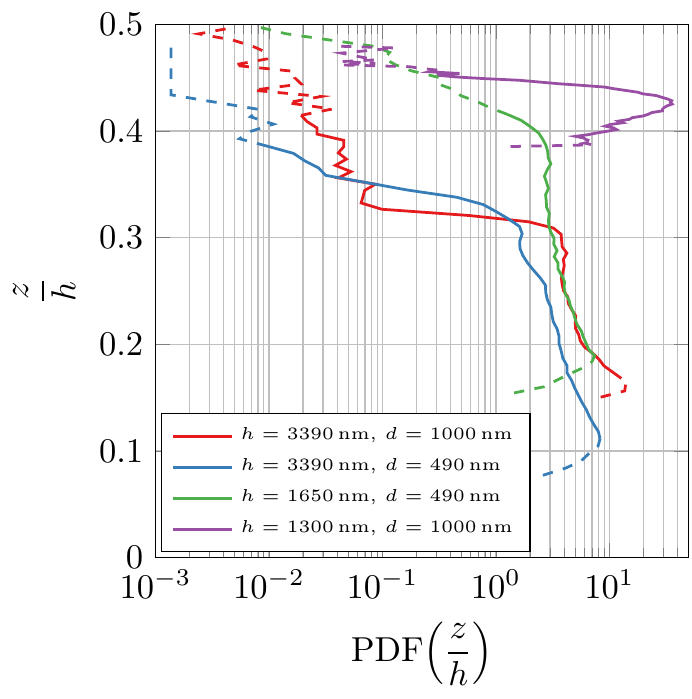}
\caption{PDF of the position $z$ of the beads in half of the nanoslits depth. Dashed parts correspond to bias-affected positions (see text for details). The center of the nanoslits correspond to $z/h=0.5$.}
\label{pdf_posz_nb}
\end{center}
\end{figure}

These observations are now used as a reference to understand the behaviours observed {for the two largest beads}. Figure \ref{pdf_nb} shows the PDF of the velocity fluctuations normalized by their standard deviation. Using this normalisation we can compare PDF for set of events with different experimental configurations (flow velocity, confinement). For each configuration $\{h,d\}$ one pressure drop is represented. The corresponding computed position PDF of the beads are plotted on figure \ref{pdf_posz_nb}. First, the two experiments in nanoslits of depth $h=3390$\,nm -- where $\beta$ is lower than 1 -- reveal a shape of the velocity PDF similar to the one observed on figure \ref{pdf_transition} for high $\Delta P$. The corresponding position PDF shows a large depletion for $z/h>0.3$ (close to the center). This confirms for this case that the gap between experimental and predicted velocities is also due to an inhomogeneous distribution of beads in the nanoslit depth{, not compensated by Brownian diffusion}. On the contrary, in the case $h=1650$\,nm with beads of diameter $d=490$\,nm, we find a velocity PDF shape similar to the ones observed in figure \ref{pdf_transition} in the ``homogeneous'' regime. This is confirmed by the position PDF which reveals a flat distribution up to $z/h=0.4$. In the last case (purple curve), a very confined configuration ($r=0.77$), the velocity PDF has a quite different shape. It is due to the narrowness of the accessible velocity range for the beads. The accessible range of position for the beads is also very thin (see figure \ref{pdf_posz_nb}). Even if beads are {weakly} Brownian in this case, the confinement is high enough to allow a homogeneous distribution of the beads in the nanoslits.

The existence of two regimes (``homogeneous'', with $\beta\approx1$ and ``inhomogeneous'', with $\beta<1$) is an {non trivial} behaviour of the transport of nano-objects in narrow channels. The explanation lies in the shape of the position distribution of the beads in the nanoslits depth. The geometry of the nanoslits entrance and the subsequent tightening of the streamlines coupled to the finite size of the beads clearly lead to a larger bead concentration close to the top wall of the nanoslits. The resulting mean velocity is intimately linked to Brownian diffusion and confinement. If either confinement or Brownian diffusion is high enough the beads can be homogenized at the pore entrance. If it is not the case, if the fluid velocity in the nanoslits is too large, the bead distribution remains inhomogeneous. The construction of an entrance P\'eclet number comparing advection and diffusion processes at the entrance of the nanoslits clearly separates these two regimes. Moreover the regime transition observed for the smallest beads at low confinement is consistent with this P\'eclet number. {Devices with similar steps are very common in microfluidic studies and applications and this effect does not seem, to our knowledge, taken into account.}

\section{Polymersomes: deviations from solid beads}

A key parameter could change the observations made above: the particle nature. In the previous section we used solid latex beads. Polymersomes as described in the subsection \ref{subsection:suspensions} are transported inside the same nanoslits as used before. It should be emphasized that these experiments are much more delicate than for commercial beads. Indeed, this type of self-assembly, usually made for micelles of typical size 20--50\,nm, does not have a perfect reproducibility for such large objects, even if the  protocol we use is optimized to get objects in the 200\,nm--1\,$\mu$m range \cite{dionzou2016}. In addition, object are only stable for around 10 days. Another key point is the polydispersity of the suspension. It can lead to clogging of the nanoslits by some objects bigger than $h$ and dramatically disturb the experiments. {However we did not observe significant adhesion of polymersomes on the walls inside the nanoslits (both zeta potential of silicon and polymersomes are negative).}

Some numerical simulations were performed to check a potential effect of polydispersity on mean velocity and PDF. We adapted the script used by Ranchon \emph{et al.} \cite{ranchon2015} which takes into account Fax\'en's law, hydrodynamics interactions and Brownian diffusion. A Gaussian size distribution with the standard deviation presented in table \ref{table:particles} was used. No significant difference with a monodisperse solution was observed (not shown here). 

\begin{figure}[h!]
\begin{center}
\includegraphics[width=0.49\textwidth]{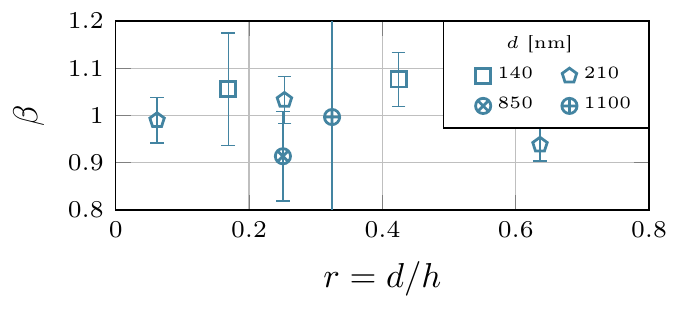}
\caption{Ratio between experimental and predicted velocity proportionality coefficients with $\Delta P$ as a function of the confinement for polymersomes. See text for details about error bars.}
\label{mean_v_poly}
\end{center}
\end{figure}

The ratio $\beta=\alpha_{exp}/\alpha_t$ is computed exactly in the same way as presented in the previous section. Figure \ref{mean_v_poly} shows this ratio as a function of confinement. 

%These experiments are rather difficult to made. First the polymersomes must be synthesised at laboratory whereas latex beads are commercial devices. Furthermore the stability of the objects dos not exceed 10 days. However it is quite difficult to choose the mean size of polymersomes with our protocol \cite{dionzou2016}.

%The large error bar is due to an experiment with few pressure drops leading to a large uncertainty in the affine fit.

%\begin{figure}[h!]
%\begin{center}
%\includegraphics[width=0.45\textwidth]{Figures/Vmoy/meanV_confinement_h_normalise_IH_rec_incert_polymersomes.pdf}
%\caption{Ratio between experimental and predicted velocity proportionality coefficients with $\Delta P$ as a function of (a) the confinement and (b) the nanoslits depth for polymersomes. See text for details about error bars. The large error bar is due to an experiment with few pressure drops leading to a large uncertainty in the affine fit. The legend is valid for both plots.}
%\label{mean_v_poly}
%\end{center}
%\end{figure}

\begin{figure}[h!]
\begin{center}
\includegraphics[width=0.49\textwidth]{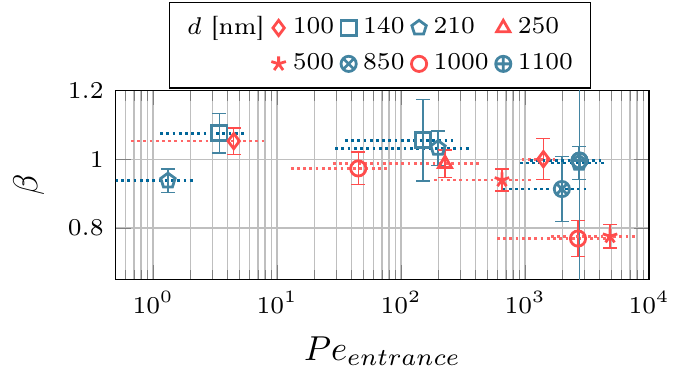}
\caption{Ratio between experimental and predicted velocity proportionality coefficients with $\Delta P$ as a function of the entrance P\'eclet number for polymersomes (blue) and a selection of beads presented figure \ref{beta_pe} (red). The dashed lines show the extension of $Pe_{entrance}$ corresponding to the different used $\Delta P$, the point is chosen to be at the center of the P\'eclet range.}
\label{beta_pe_poly}
\end{center}
\end{figure}

Contrary to the beads we do not clearly observe two regimes. {For all the different polymersomes diameter}, the experimental mean velocity does not really differ from the expected $V_t$. To compare to beads, $\beta$ is plotted as a function of $Pe_{entrance}$ (eq. \ref{eq:Pe}) on figure \ref{beta_pe_poly}, together with a selection of typical results for beads (selected from figure \ref{beta_pe}). Polymersomes spread within a similar range of P\'eclet number as beads: we have both experiments at high confinement and Brownian diffusion, and experiments at low confinement and Brownian diffusion. It seems to confirm that there is {no} a net appearance of the ``inhomogeneous'' regime with polymersomes: for $Pe_{entrance}>10^3$ the polymersomes reach $\beta\approx1$ whereas beads remains to $\beta<0.8$. Nevertheless polymersomes at high $Pe_{entrance}$ could be in the transition between the two regimes, with a value of $\beta$ slightly lower than unity, but still at the limit of resolution due to error bars.

\begin{figure}[h!]
\begin{center}
\includegraphics[width=0.45\textwidth]{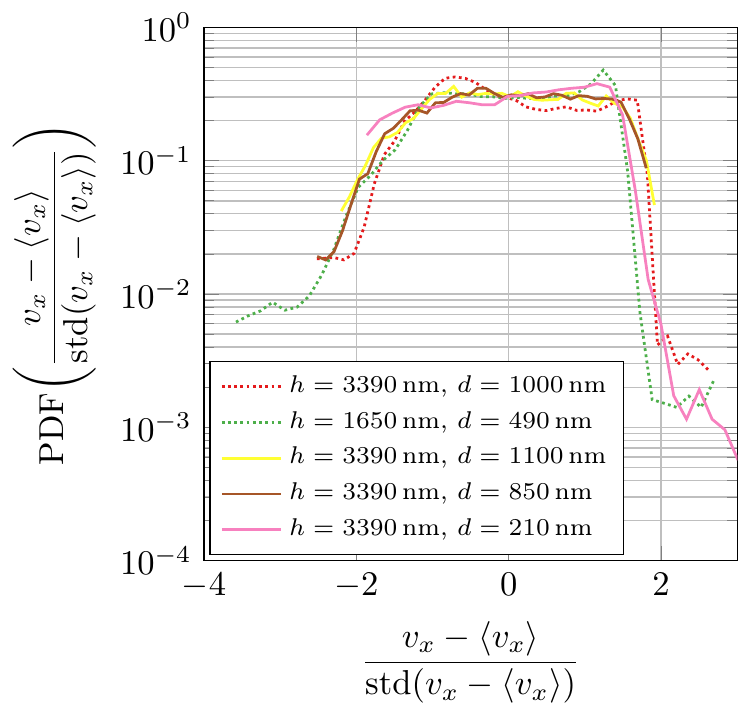}
\caption{PDF of the longitudinal velocity $v_x$ in different configurations for polymersomes (solid lines) and a selection of experiments presented figure \ref{pdf_nb} (dashed).}
\label{pdf_poly}
\end{center}
\end{figure}

\begin{figure}[h!]
\begin{center}
\includegraphics[width=0.45\textwidth]{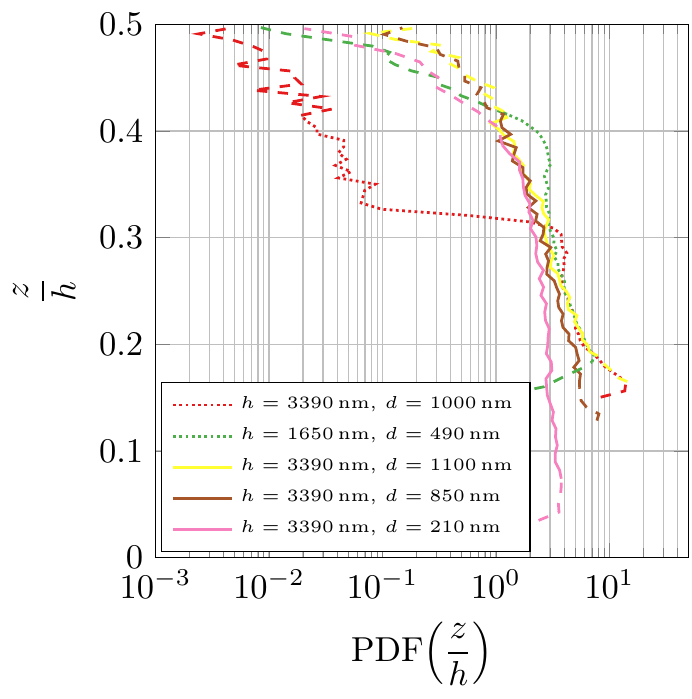}
\caption{PDF of the particles position $z$ in the nanoslits depth. Different configurations for polymersomes (thick lines) and a selection of beads presented figure \ref{pdf_nb} (thin lines) are shown. Dashed parts correspond to bias-affected positions (see text for details). The center of the nanoslits correspond to $z/h=0.5$.}
\label{pdf_pos_poly}
\end{center}
\end{figure}

In order to go further in this analysis, some velocity PDF are plotted on figure \ref{pdf_poly} and compared to two velocity PDF for beads taken from both regimes. Contrary to the beads in the ``inhomogeneous'' regime we do not clearly observe, for polymersomes, PDF with a tail higher than the high-velocity peak (we remind here that this tail in the second regime is attributed to an extra-concentration of beads at low velocities, i.e. close to the wall). However this high-velocity peak is not really visible on the PDF for polymersomes of mean diameter $d=850$ and 1100\,nm (high $Pe_{entrance}$ values): these configurations reveal a flatter PDF than for beads at similar $Pe_{entrance}$ values ($h=3390$\,nm, $d=1000$\,nm). For the case of highly Brownian polymersomes and low confinement ($h=3390$\,nm, $d=210$\,nm, low $Pe_{entrance}$ value), the velocity PDF shows more probability of high velocities, even if it is less marked than beads in the ``homogeneous'' regime. These different observations tend to confirm that we could have a transition to the ``inhomogeneous'' regime for polymersomes in configurations with {weakly} Brownian polymersomes and low confinement. The corresponding $z$ position PDF presented in figure \ref{pdf_pos_poly} seem to confirm this assertion. In the case of highly Brownian polymersomes and low confinement ($h=3390$\,nm, $d=210$\,nm), the position PDF is flat, similarly to the one observed for beads of diameter $d=490$\,nm in nanoslits of depth $h=1650$\,nm: these objects are distributed homogeneously in the nanoslits depth. In the case of {weakly} Brownian polymersomes and low confinement ($h=3390$\,nm, $d=850$, 1100\,nm, high $Pe_{entrance}$ values) the position PDF are very different from the beads one in the ``inhomogeneous'' regime ($h=3390$\,nm, $d=1000$\,nm), but they also differ from the beads one in the ``homogeneous'' regime: they do not reveal a flat distribution. %Actually their shape seems quite similar to the one we observed on figure \ref{pdf_posz_transition} at high $\Delta P$. 
%In the largely confined case, the PDF is similar to the purple one ($h=1300$\,nm, $d=1000$\,nm) visible in figure \ref{pdf_nb}.

Actually the shape of velocity PDF for {weakly} Brownian polymersomes at low confinement seems quite similar to the one we observed in figure \ref{pdf_posz_transition} at high $\Delta P$: the largest polymersomes in the deepest nanoslits seem to be transitioning between the two regimes. The capsule shape of the polymersomes could shift the transition between the two regimes towards larger P\'eclet numbers. At pore entrance scale, because of the step between the microchannel and the nanoslits, polymersomes could have a different trajectory and thus be transported on different streamlines into the nanoslits. It would result in a reduction of the distribution inhomogeneity when confinement and Brownian diffusion are both low (less marked). This difference could be explained  by the softness and shape of these objects, according to several mechanisms: change of the shape of the polymersomes at the entrance of the nanoslits, viscous dissipation at the surface of the capsule or flow inside it. Zhu \emph{et al.} \cite{zhu2014} recently observed using numerical simulations that the way a soft capsule transported in a flow skirts an obstacle dramatically depends on the softness of this capsule. {Since the characterization of the mechanical properties of polymersomes and of their behaviour under flow are really challenging, a further interpretation is delicate.}

%The competition between viscous and elastic forces undergone by the polymersomes can be represented by a capillary number: 

%\begin{equation}
%Ca=\frac{\eta V_t}{G_s},
%\label{ca}
%\end{equation}
%
%\noindent where $G_s$ is the isotropic shear modulus \cite{zhu2014}. In our case, assuming a typical velocity $V_t$ of 1\,mm.s$^{-1}$, the capillary number can be approximated by $Ca=XX$.  TO BE CONTINUED

\section{Conclusion}

We study experimentally the transport of rigid, solid beads and soft, capsule-shaped polymersomes in silicon-etched micrometric and sub-micrometric channels. Three parameters {and their cross-effects} are investigated: Brownian diffusion, confinement and object nature. Two main regimes are observed, corresponding to a transition from homogeneous to inhomogeneous particle distribution in the nanoslits depth. Respectively, the mean experimental velocity is comparable to or 20\% lower than the predicted one. The first regime occurs when the Brownian diffusion is able to homogenize the particles at the pore entrance, either because of the small particle size or of the high confinement. The second one appears at low Brownian diffusion and low confinement. An entrance P\'eclet number, which compares advection and diffusion at the pore entrance, shows a clear distinction between these two regimes. The transition between these two regimes is observed with a very low confinement ($r=0.03$) and highly Brownian  beads by varying the advection velocity. It is related to a dramatic change in the velocity distribution change: low velocities become more probable than high ones when the pressure drop which controls the advection velocity rises. The deduction of the beads position distribution in the nanoslits from the velocity distributions clearly shows this transition from a homogeneous to an inhomogeneous repartition of the objects. This deduction confirms that the two regimes observed on mean velocities are directly linked to this difference of distribution. Such a distribution could be driven by the pore entrance geometry. The step between the microchannel and the nanoslits combined to the finite size of the particles forces to cross streamlines and to concentrate close to the top wall of the nanoslits. The capsule shape of polymersomes seems to shift the onset of the second regime. Whereas the mean experimental velocity for the biggest polymersomes seems close to the expected one, the velocity and position distributions show a behaviour similar to the transition state evoked above. This shift could be due to a trajectory at the pore entrance quite different for such objects. 

In conclusion, confined transport of colloids is quantitatively described by a hydrodynamic model (Fax\'en averaging of the velocity field and hydrodynamic interaction with the wall), even at high confinement. However we demonstrate an entrance effect, reminiscent of hydrodynamic filtration, should be relevant  in many configurations involving confined flows: it is related to the entry of particles from a deep microchannel or reservoir to a narrower channel. Finally this reveals the importance to take into account both Brownian diffusion, particle nature and confinement when microfluidic systems are used to manipulate nano-objects, particularly biological or biomimetics objects. {An exciting perspective could be to characterize precisely the behaviour of polymersomes under flow, and to relate it to their mechanical properties, in order to understand the difference observed on the transport in narrow channels from solid, rigid beads.}

\section*{Acknowledgements}

  We acknowledge the F\'ed\'eration FERMaT, University of Toulouse (Project NEMESIS) and ANR (Project PolyTransFlow, contract n\textsuperscript{o}ANR-13-BS09-0015) for funding these researches. O. Liot and P. Joseph warmly acknowledge H. Tabuteau, C. Roux and A. Bancaud for fruitful discussions. This work was partly supported by LAAS-CNRS micro and nanotechnologies platform member of the French RENATECH network. \\

\bibliographystyle{Vancouver}
\bibliography{biblio}

\begin{thebibliography}{10}

\bibitem{ranchon2016}
Ranchon H, Malbec R, Picot V, Boutonnet A, Terrapanich P, Joseph P, et~al.
\newblock {DNA} separation and enrichment using electro-hydrodynamic
  bidirectional flows in viscoelastic liquids.
\newblock Lab on a Chip. 2016;16(7):1243--1253.
\newblock Available from:
  \url{http://pubs.rsc.org/en/Content/ArticleLanding/2016/LC/C5LC01465D}.

\bibitem{bradford2008}
Bradford SA, Torkzaban S.
\newblock Colloid {Transport} and {Retention} in {Unsaturated} {Porous}
  {Media}: {A} {Review} of {Interface}-, {Collector}-, and {Pore}-{Scale}
  {Processes} and {Models}.
\newblock Vadose Zone Journal. 2008;7(2):667.
\newblock Available from:
  \url{https://www.soils.org/publications/vzj/abstracts/7/2/667}.

\bibitem{wyss2006}
Wyss HM, Blair DL, Morris JF, Stone HA, Weitz DA.
\newblock Mechanism for clogging of microchannels.
\newblock Physical Review E. 2006 Dec;74(061402).
\newblock Available from:
  \url{https://link.aps.org/doi/10.1103/PhysRevE.74.061402}.

\bibitem{dressaire2017}
Dressaire E, Sauret A.
\newblock Clogging of microfluidic systems.
\newblock Soft Matter. 2017;13(1):37--48.
\newblock Available from: \url{http://xlink.rsc.org/?DOI=C6SM01879C}.

\bibitem{wu2009}
Wu Z, Hjort K.
\newblock Microfluidic {Hydrodynamic} {Cell} {Separation}: {A} {Review}.
\newblock Micro and Nanosystems. 2009 Nov;1(3):181--192.

\bibitem{zhang2012}
Zhang W, Tang X, Weisbrod N, Guan Z.
\newblock A review of colloid transport in fractured rocks.
\newblock Journal of Mountain Science. 2012 Dec;9(6):770--787.
\newblock Available from:
  \url{http://link.springer.com/10.1007/s11629-012-2443-1}.

\bibitem{staben2003}
Staben ME, Zinchenko AZ, Davis RH.
\newblock Motion of a particle between two parallel plane walls in
  low-{Reynolds}-number {Poiseuille} flow.
\newblock Physics of Fluids. 2003;15(6):1711.
\newblock Available from:
  \url{http://scitation.aip.org/content/aip/journal/pof2/15/6/10.1063/1.1568341}.

\bibitem{staben2005}
Staben ME, Davis RH.
\newblock Particle transport in {Poiseuille} flow in narrow channels.
\newblock International Journal of Multiphase Flow. 2005 May;31(5):529--547.
\newblock Available from:
  \url{http://linkinghub.elsevier.com/retrieve/pii/S0301932205000066}.

\bibitem{pasol2011}
Pasol L, Martin M, Ekiel-Jezewska ML, Wajnryb E, Blawzdziewicz J, Feuillebois
  F.
\newblock Motion of a sphere parallel to plane walls in a {Poiseuille} flow.
  {Application} to field-flow fractionation and hydrodynamic chromatography.
\newblock Chemical Engineering Science. 2011 Sep;66(18):4078--4089.
\newblock Available from:
  \url{http://linkinghub.elsevier.com/retrieve/pii/S0009250911003423}.

\bibitem{ranchon2015}
Ranchon H, Picot V, Bancaud A.
\newblock Metrology of confined flows using wide field nanoparticle
  velocimetry.
\newblock Scientific Reports. 2015 Sep;5(1).
\newblock Available from: \url{http://www.nature.com/articles/srep10128}.

\bibitem{lu2017}
Lu X, Liu C, Hu G, Xuan X.
\newblock Particle manipulations in non-{Newtonian} microfluidics: {A} review.
\newblock Journal of Colloid and Interface Science. 2017 Aug;500:182--201.
\newblock Available from:
  \url{http://linkinghub.elsevier.com/retrieve/pii/S0021979717304113}.

\bibitem{charru2007}
Charru F, Larrieu E, Dupont JB, Zenit R.
\newblock Motion of a particle near a rough wall in a viscous shear flow.
\newblock Journal of Fluid Mechanics. 2007 Jan;570:431.
\newblock Available from:
  \url{http://www.journals.cambridge.org/abstract_S0022112006003090}.

\bibitem{ranchon2018}
Ranchon H, Cacheux J, Reig B, Liot O, Terrapanich P, Leichl{\'e} T, et~al.
\newblock Accelerated transport of particles in confined channels with high
  roughness amplitude.
\newblock Langmuir. 2018 Jan;Available from:
  \url{http://dx.doi.org/10.1021/acs.langmuir.7b03962}.

\bibitem{urzay2007}
Urzay J, Llewellyn~Smith SG, Glover BJ.
\newblock The elastohydrodynamic force on a sphere near a soft wall.
\newblock Physics of Fluids. 2007 Oct;19(10):103106.
\newblock Available from: \url{http://aip.scitation.org/doi/10.1063/1.2799148}.

\bibitem{davies2017}
Davies H, D{\'e}barre D, Verdier C, Richter RP, Bureau L.
\newblock Lift at a soft wall; 2017.
\newblock Available from: \url{https://hal.archives-ouvertes.fr/hal-01652253}.

\bibitem{maeda2012}
Maeda H.
\newblock Macromolecular therapeutics in cancer treatment: the {EPR} effect and
  beyond.
\newblock Journal of Controlled Release: Official Journal of the Controlled
  Release Society. 2012 Dec;164(2):138--144.

\bibitem{kumar2012}
Kumar A, Graham MD.
\newblock Margination and segregation in confined flows of blood and other
  multicomponent suspensions.
\newblock Soft Matter. 2012 Oct;8(41):10536--10548.
\newblock Available from:
  \url{http://pubs.rsc.org/en/content/articlelanding/2012/sm/c2sm25943e}.

\bibitem{dionzou2016}
Dionzou M, Mor{\`e}re A, Roux C, Lonetti B, Marty JD, Mingotaud C, et~al.
\newblock Comparison of methods for the fabrication and the characterization of
  polymer self-assemblies: what are the important parameters?
\newblock Soft Matter. 2016 Feb;12(7):2166--2176.

\bibitem{lefebvre2007}
Lefebvre Y, Barth{\`e}s-Biesel D.
\newblock Motion of a capsule in a cylindrical tube: effect of membrane
  pre-stress.
\newblock Journal of Fluid Mechanics. 2007 Oct;589:157--181.
\newblock Available from:
  \url{https://www.cambridge.org/core/journals/journal-of-fluid-mechanics/article/motion-of-a-capsule-in-a-cylindrical-tube-effect-of-membrane-pre-stress/D7FEEAA036305F792F68C49113FA895B}.

\bibitem{vlahovska2009}
Vlahovska PM, Podgorski T, Misbah C.
\newblock Vesicles and red blood cells in flow: {From} individual dynamics to
  rheology.
\newblock Comptes Rendus Physique. 2009 Nov;10(8):775--789.
\newblock Available from:
  \url{https://www.scholars.northwestern.edu/en/publications/vesicles-and-red-blood-cells-in-flow-from-individual-dynamics-to-}.

\bibitem{stetefeld2016}
Stetefeld J, McKenna SA, Patel TR.
\newblock Dynamic light scattering: a practical guide and applications in
  biomedical sciences.
\newblock Biophysical Reviews. 2016 Oct;8(4):409--427.
\newblock Available from:
  \url{https://www.ncbi.nlm.nih.gov/pmc/articles/PMC5425802/}.

\bibitem{le_meins2011}
Le~Meins JF, Sandre O, Lecommandoux S.
\newblock Recent trends in the tuning of polymersomes{\textquoteright} membrane
  properties.
\newblock The European Physical Journal E. 2011 Feb;34(2).
\newblock Available from:
  \url{http://link.springer.com/10.1140/epje/i2011-11014-y}.

\bibitem{jaskiewicz2012}
Jaskiewicz K, Makowski M, Kappl M, Landfester K, Kroeger A.
\newblock Mechanical {Properties} of
  {Poly}(dimethylsiloxane)-block-poly(2-methyloxazoline) {Polymersomes}
  {Probed} by {Atomic} {Force} {Microscopy}.
\newblock Langmuir. 2012 Aug;28(34):12629--12636.
\newblock Available from: \url{http://dx.doi.org/10.1021/la301608k}.

\bibitem{kirby2004}
Kirby BJ, Hasselbrink EF.
\newblock Zeta potential of microfluidic substrates: 1. {Theory}, experimental
  techniques, and effects on separations.
\newblock Electrophoresis. 2004 Jan;25(2):187--202.

\bibitem{naillon2016}
Naillon A.
\newblock {\'E}coulements liquide-gaz, {\'e}vaporation, cristallisation dans
  les milieux micro et nanoporeux: {\'e}tudes {\`a} partir de syst{\`e}mes
  mod{\`e}les micro et nanofluidiques.
\newblock Universit{\'e} Paul Sabatier. Toulouse; 2016.

\bibitem{chen2000}
Chen, Ye.
\newblock Faxen's {Laws} of a {Composite} {Sphere} under {Creeping} {Flow}
  {Conditions}.
\newblock Journal of Colloid and Interface Science. 2000 Jan;221(1):50--57.

\bibitem{bruus2007}
Bruus H.
\newblock Theoretical microfluidics.
\newblock Oxford university press Oxford; 2007.
\newblock Available from:
  \url{http://web-files.ait.dtu.dk/bruus/TMF/publications/books/Bruus_TMFbook_Sample_Chapter.pdf}.

\bibitem{bhattacharya2005}
Bhattacharya S, B{\l }awzdziewicz J, Wajnryb E.
\newblock Hydrodynamic interactions of spherical particles in suspensions
  confined between two planar walls.
\newblock Journal of Fluid Mechanics. 2005 Oct;541:263--292.
\newblock Available from:
  \url{https://www.cambridge.org/core/journals/journal-of-fluid-mechanics/article/hydrodynamic-interactions-of-spherical-particles-in-suspensions-confined-between-two-planar-walls/41606F43C15AB5687562DFC8DC07B9C0}.

\bibitem{mcgrath2014}
McGrath J, Jimenez M, Bridle H.
\newblock Deterministic lateral displacement for particle separation: a review.
\newblock Lab Chip. 2014;14(21):4139--4158.
\newblock Available from: \url{http://xlink.rsc.org/?DOI=C4LC00939H}.

\bibitem{yamada2004}
Yamada M, Nakashima M, Seki M.
\newblock Pinched {Flow} {Fractionation}: {Continuous} {Size} {Separation} of
  {Particles} {Utilizing} a {Laminar} {Flow} {Profile} in a {Pinched}
  {Microchannel}.
\newblock Analytical Chemistry. 2004 Sep;76(18):5465--5471.
\newblock Available from: \url{http://dx.doi.org/10.1021/ac049863r}.

\bibitem{dersoir2015-1}
Dersoir B.
\newblock La physique du colmatage : de la particule collo{\"i}dale au bouchon.
\newblock Universit{\'e} de Rennes. Rennes; 2015.
\newblock Available from:
  \url{https://hal.archives-ouvertes.fr/tel-01188553/document}.

\bibitem{vasseur1976}
Vasseur P, Cox RG.
\newblock The lateral migration of a spherical particle in two-dimensional
  shear flows.
\newblock Journal of Fluid Mechanics. 1976 Nov;78(02):385.
\newblock Available from:
  \url{http://www.journals.cambridge.org/abstract_S0022112076002498}.

\bibitem{schonberg1989}
Schonberg JA, Hinch EJ.
\newblock Inertial migration of a sphere in {Poiseuille} flow.
\newblock Journal of Fluid Mechanics. 1989;203:517--524.

\bibitem{dersoir2017}
Dersoir B, Schofield AB, Tabuteau H.
\newblock Clogging transition induced by self filtration in a slit pore.
\newblock Soft Matter. 2017;13(10):2054--2066.
\newblock Available from: \url{http://xlink.rsc.org/?DOI=C6SM02605B}.

\bibitem{fouet2016}
Fouet M, Mader MA, Ira{\"i}n S, Yanha Z, Naillon A, Cargou S, et~al.
\newblock Filter-less submicron hydrodynamic size sorting.
\newblock Lab on a Chip. 2016;16(4):720--733.
\newblock Available from:
  \url{http://pubs.rsc.org/en/Content/ArticleLanding/2016/LC/C5LC00941C}.

\bibitem{zhu2014}
Zhu L, Rorai C, Dhrubaditya~Mitra DM, Brandt L.
\newblock A microfluidic device to sort capsules by deformability: a numerical
  study.
\newblock Soft Matter. 2014;10(39):7705--7711.
\newblock Available from: \url{http://xlink.rsc.org/?DOI=C4SM01097C}.

\end{thebibliography}
\end{document}